\documentclass[aps,prd,onecolumn,showkeys,groupedaddress,showpacs,nofootinbib]{revtex4}
\usepackage{epsfig} \usepackage{amsmath} \usepackage{amsfonts}
\usepackage{amssymb} \usepackage{graphicx} \usepackage{colordvi}
\usepackage{psfrag}
 \usepackage{times}
 \usepackage{amsmath, amsthm, amssymb}
 \usepackage{makeidx}

\newfont{\gotico}{eufm10 scaled\magstephalf}
\newfont{\qvd}{msam10 scaled\magstephalf}

\def\de#1/de#2{\frac{\partial {#1}}{\partial {#2}}}
\def\De#1/de#2{\dfrac{\partial {#1}}{\partial {#2}}}

\makeindex \makeatletter
\def\widebar{\accentset{{\cc@style\underline{\mskip10mu}}}}
\makeatother

\begin{document}
\def\bib#1{[{\ref{#1}}]}
\title{\bf A tomographic description for classical and quantum cosmological perturbations}

\author{S. Capozziello$^{1,2}$, V.~I.~Man'ko$^{1,2,3}$, G.~Marmo$^{1,2}$ and C.~Stornaiolo$^{2}$}

\affiliation{$^{1}$ Dipartimento di Scienze Fisiche,
Universit\`{a} ``Federico II'' di Napoli and $^{2}$ INFN Sez. di
Napoli, Compl. Univ. Monte S. Angelo Ed. N, via Cinthia, I- 80126
Napoli (Italy)}

\affiliation{$^{3}$ P.N. Lebedev Physical Institute, Leninskii Pr.
53, Moscow 117924 (Russia).}

\date{\today}

\begin{abstract}
Classical and quantum  perturbations can be described in terms of
marginal distribution functions in the framework of tomographic
cosmology. In particular, the so called Radon transformation and
the mode-parametric quantum oscillator description can give rise
to links between quantum and classical regimes. The approach
results a natural scheme to discuss the transition from the
quantum  to the classical perturbations and then it could be a
workable scheme to connect primordial fluctuations with the today
observed large scale structure.

\end{abstract}


\maketitle

\vskip 0.5truecm
\section{Introduction}\label{introduction}

Addressing the problem of  cosmic evolution, starting from the
inflationary mechanism  \cite{Guth:1980zm,Linde:1981mu} up to the
observed scenario, as well as the problem of producing classical
perturbations starting from the quantum ones, have been the
subject of several studies  during the last decades
\cite{Mukhanov:1990me,Lesgourgues:1996jc,Kiefer:1998qe}. In
particular, the Friedmann-Lemaitre-Robertson-Walker (FLRW)
homogeneous and isotropic metric,  described by the expansion
factor $a(t)$, is usually considered  the standard background on
which formulate the classical and quantum theory of cosmological
perturbations. The goal is to provide a suitable description
which, starting from quantum regime can lead  to the observed
large-scale structures such as galaxies, galaxy clusters,
super-clusters and voids.

On the other hand, the task of quantum  cosmology is to achieve a
quantum description of the cosmic initial state which, by the
following evolution could determine, in principle, all the
features of the latest classical epochs which we today observe by
the astrophysical measurements.  The initial quantum  state is
usually associated with the so-called "Wave Function of the
Universe", in analogy with standard quantum mechanics
\cite{schroedinger26}, or with its  density matrix
\cite{landau,vonneumann32}. A useful representation to consider
the quantum-to-classical transition for the quantum  state of the
universe is the Wigner function $W(q,p)$ (see, e.g.,
\cite{Kiefer:1998qe,hl}).

Such a Wigner function is the analogue of the classical
probability distributions on phase-space $f(p,q)$ which are, in
general,  non-negatively defined. On the other hand, due to the
uncertainty relations, the Wigner function can assume negative
values in some domain of phase-space. In view of this fact, it is
not a fair probability distribution and it is called a
quasi-distribution.

Besides, quantum cosmology in its development has always adopted
new results from quantum mechanics and quantum field theory.
Recently, a new tomographic probability representation of quantum
states  has been  found \cite{Mancini:1996wd}.  In this picture,
the fair nonnegative probability distribution, containing a
complete information on the states, is used instead of the wave
function or the density matrix. Though the new probability
representation is essentially equivalent to all the other
available representations adopted in quantum mechanics
{\cite{9qm}, it has its own merits. An important property of this
representation is that the quantum state   is associated to the
same tomographic probability density $\mathcal{W}(X,\mu,\nu)$,
connected with the Wigner function by the standard integral Radon
transform \cite{radon}. Besides,  the classical state is described
 using the Radon component of the standard classical
probability density $f(p,q)$ \cite{olga97}. Thus, in the
tomographic probability representation, both classical and quantum
states can be described by the same non-negative probability
densities $\mathcal{W}(X,\mu,\nu)$: this fact makes easier to
consider the classical-to-quantum mutual relations and
transitions.

Some aspects of quantum cosmology have already been studied in the
framework of the tomographic probability approach giving a picture
coherent with both inflationary and standard Friedmann cosmology
\cite{Manko:2003dp,Man'ko:2004zj,Man'ko:2006wv,Stornaiolo:2006da,Stornaiolo:2006kz,Capozziello:2007xn}.
Since  the issue of quantum-to-classical transition  plays a main
important role in the theory of cosmological perturbations, the
aim of this paper is to study the perturbations using the
tomographic probability densities for the description of
cosmological "states" (both classical and quantum ones). Such
"states" are configurations of the universe which can be suitably
defined by assigning a set of cosmological observables. As shown
in \cite{Mukhanov:1990me}, it is possible to find the same
equations for perturbations in the cases of  hydrodynamic matter,
of  scalar field and of  alternative  theories of gravity as
$f(R)$. The perturbation evolution can be described by the
Hamiltonian formalism corresponding to small vibrations on a
classical background. The field theory can be considered in the
tomographic picture as shown in \cite{rosapl}. The Hamiltonian for
small vibrations  corresponds to interacting oscillators with
time-dependent frequencies and mutual coupling constants. In
quantum mechanics, the one-dimensional parametric oscillator has
been studied by Husimi \cite{hus1953} and new integrals of motion,
linear in field quadratures  of the oscillator have been found in
\cite{Malkin:1970gq,Malkin:1971gs}. A comprehensive approach to
the system of interacting parametric oscillators has been
presented in \cite{markov}. The theory of such oscillators with
dissipation can be developed in connection with the non-stationary
Casimir effect studied in \cite{DodDod}.

The advantage of the  tomographic approach for this particular
Hamiltonian is related to the fact that,  in both quantum and
classical domains, the propagators providing the tomograms of
states at time $t$, in terms of the states tomograms at some
earlier time, are identical. For these particular propagators, the
difference between the quantum and classical universe descriptions
is connected to the choice of the initial state tomogram only.

The layout of the paper is the following. In Sect. II,  the main
points of the field theory for cosmological perturbations is
sketched.  The probability representation of cosmological
perturbations is discussed in Sect.III. A general scheme for
approaching classical and quantum perturbations is given in Sect
IV.  Sect.V is devoted to  a detailed description of the one-mode
parametric quantum oscillator, while the multi-mode case is
considered in Sect.VI in view of the classical-to-quantum
relations. Conclusions are drawn in Sec.VII.

\section{The field theory of perturbations}

In \cite{Mukhanov:1990me}, a complete gauge invariant theory for
the cosmological perturbations is discussed. The cases of
hydrodynamic and scalar fields are studied and the formulation in
terms of theories alternative to general relativity is considered.

Gauge invariance is a crucial requirement for quantization,
because it allows to properly quantize  the physical degrees of
freedom.

According to the inflationary paradigm, the origin of the
cosmological perturbations, which lead to the hierarchy of
structures which form the observed universe, can be retraced back to
the quantum fluctuations of the very early universe.

The quantization of cosmological perturbations allows, in
principle, to study the initial state of the universe, at least
from a theoretical point of view. In order to relate the early
quantum states of the universe to the cosmological observations,
it is necessary to develop a model for the evolution of the
quantum fluctuations into classical perturbations.

A unifying picture of these different models for perturbations has
been proposed in order to quantize them.

To this aim, a total action for gravity and matter is
\begin{equation}\label{gravaction}
    S= \frac{1}{16\pi G}\int R \sqrt{-g} d^{4}x +
    \int\mathcal{L}_{m}(g) \sqrt{-g} d^{4}x
\end{equation}
if we are considering General Relativity. On the other hand,
actions of the forms

\begin{equation}\label{alternativegravaction}
    S= \int f(R) \sqrt{-g} d^{4}x +
    \int\mathcal{L}_{m}(g) \sqrt{-g} d^{4}x\,,
\end{equation}

 \begin{equation} \label{scalaraction}
 S=\int
 \sqrt{-g}d^{4}x\left[F(\Phi)R+\frac{1}{2}g^{\mu\nu}\Phi_{;\mu}\phi_{;\,\nu}
 -W(\Phi)\right]+\int\mathcal{L}_{m}(g) \sqrt{-g} d^{4}x\,,
 \end{equation}

where $f(R), F(\Phi), W(\Phi)$ are generic functions of the Ricci
scalar $R$ and a scalar field $\Phi$, respectively can be adopted
in view to face several cosmological problems ranging from
inflation to dark energy (see e.g.
\cite{copeland,odintsov,faraoni,GRGrew} for reviews).

Considering the simplest case (General Relativity), the action for
the perturbations can be derived from (\ref{gravaction}) by
writing first the metric (in the conformal time $\eta$) in the ADM
formulation
\begin{equation}\label{ADMmetric}
  ds^{2}=(\mathcal{N}^{\,2}-\mathcal{N}_{i}\mathcal{N}^{\,i})d\eta^{2}
  -2 \mathcal{N}_{i}dx^{i}d\eta -\gamma_{ij}dx^{i}dx^{j}\,,
\end{equation}

where the lapse function is
\begin{equation}\label{lapse}
    \mathcal{N}=a(\eta)(1+\phi-\frac{1}{2}\phi^{2}+
    \frac{1}{2}B_{,i}B_{,i})\,,
    \end{equation}
the shift functions are
\begin{equation}\label{shifts}
\mathcal{N}_{i}=a^{2}(\eta)B_{,i}
\end{equation}
the spatial metric is
\begin{equation}\label{spatialmetric}
    \gamma_{ij}=a^{2}(1-2\psi)\delta_{ij}+2a^{2} E_{,ij}\,,
\end{equation}
and its inverse is
\begin{equation}\label{inversespatialmetric}
     \gamma^{ij}=a^{-2}(\delta_{ij}+2\psi\delta_{ij}- 2E_{,ij}+4 E_{,il}E_{lj}-8E_{ij}\psi)
\end{equation}
where   the convention of summation has been adopted on the
repeated lower indices.

The fields $\phi$, $\psi$, $E$ and $B$, introduced in the previous
equations, form the tensor $\delta g_{\mu\nu}$ which represents
the scalar perturbations of the FLRW spacetime  metric. Each of
these perturbations terms are gauge dependent, but as shown in
\cite{Mukhanov:1990me}, linear combinations of these functions and
of their derivatives can be taken into account  to construct gauge
invariant objects.

Writing the gravitational part of the action in the ADM form and
and expressing it in terms of the perturbed metric,  we arrive at
the equation  for a gauge invariant combination of the fields
$\phi$, $\psi$, $E$ and $B$. The perturbations have been studied
in three cases, i.e. hydrodynamical, scalar field and
$f(R)$-gravity. In all these three cases, the action for the
perturbations, described in terms of gauge invariant fields
(omitting total derivatives), takes the  form
\begin{equation}\label{actionfor perturbation}
   S_{perturb.}=\frac{1}{2}\int\left( {v'}^{2} -c^{2}_{s}
   \gamma^{ij}v_{,i}v_{,j}+\frac{z''}{z}v^{2}\right)\sqrt{\gamma}\,d^{4}x,
\end{equation}
where $z$ is a time-dependent function. Eq. (\ref{actionfor
perturbation})  is the action of a scalar field $v$ with  a
time-dependent mass $ {\displaystyle m^{2}(\eta)=\frac{z''}{z}}$.

Turning to the Hamiltonian  formulation, the conjugate momentum for
the field $v$ is
\begin{equation}\label{conjmoment}
     \pi(\eta,\mathbf{x})=\frac{\delta \mathcal{L}}{\delta
     v'}=v'(\eta,\mathbf{x})\,,
\end{equation}
and the resulting Hamiltonian is
\begin{equation}\label{hamiltonian}
    \mathcal{H}=\frac{1}{2}\int\left( \pi^{2} +c^{2}_{s}
    \gamma^{ij}v_{,i}v_{,j}-\frac{z''}{z}v^{2}\right)\sqrt{\gamma}\,d^{3}x\,,
\end{equation}
where $c_{s}$ is the  sound speed and
\begin{equation}\label{zeta}
    z=\frac{a(\mathcal{K}+\mathcal{H}^{2}-\mathcal{H}')^{1/2}}{\mathcal{H}c_{s}}\,.
\end{equation}
The transition to the quantum formulation is obtained once the
variables $v$ and $\pi$ are replaced with the operators $ \hat{v}$
and $\hat{\pi}$, satisfying the following commutation relations
\begin{equation}\label{commutationrelations}
     [\hat{v}(\eta,\mathbf{x}),\hat{v}(\eta,\mathbf{x'})]=[\hat{\pi}(\eta,\mathbf{x}),\hat{\pi}(\eta,\mathbf{x'})]=0,
      \ \ \ \ \ \ \ \ [\hat{v}(\eta,\mathbf{x}),\hat{\pi}(\eta,\mathbf{x'})]=i\delta(\mathbf{x}-\mathbf{x'})
\end{equation}
where the delta function $\delta(\mathbf{x}-\mathbf{x'})$ is
normalized by requiring
\begin{equation}\label{delta}
    \int\sqrt{\gamma}\delta(\mathbf{x}-\mathbf{x'})d^{3}x=1.
\end{equation}

\section{The probability representation for cosmological perturbations}

Instead of going into the usual procedure of canonical
quantization of the above system, we want to indicate a different
approach to quantize the perturbation field in terms of
probability distribution functions in the way proposed in
\cite{Mancini:1996wd}.

In this formulation of quantum mechanics,  marginal distribution
functions with classical-like evolution replace the wave
functions. The main advantage of this formulation is that  a
quantum theory can be entirely expressed  in terms of observable
functions, which are comparable with their classical counterparts.

This approach has been already applied in quantum cosmology with
the purpose to study the evolution of a quantum universe into a
classical one and to obtain all the information of the initial
quantum cosmological stages from the today observations.

Recently this approach has been extended to quantum field theory
\cite{rosapl}. We shall briefly recall the results here.

Let us consider the quantum Hamiltonian for a scalar field in a
$(d+1)$ spacetime
\begin{equation}\label{auntumhamiltonian}
    \hat{H}=\int\left[\frac{1}{2}\hat{\pi}^{2}+\frac{1}{2}\sum_{b=1}^{d}(\partial_{b}\hat{\varphi}(x))^{2}+
    U(\hat{\varphi}(x))\right]d^{d}x\,,
\end{equation}
and  the combination
\begin{equation}\label{combination}
     \hat{\Phi}(x)=\mu(x)\hat{\varphi}(x)+\nu(x)\hat{\pi}(x),
\end{equation}
introducing the quantum characteristic function\footnote{Note that
the meaning of the function $\hat{\Phi}(x)$ is different with
respect to the scalar field mentioned in the action
(\ref{scalaraction}).}
\begin{equation}\label{quantumcharacter}
     \chi(k(x))=\left\langle\ \exp \left(i\int d^{d}x  k(x)\hat{\Phi}(x)\right)\right\rangle
\end{equation}

we can define the marginal distribution functional

\begin{equation}\label{mdf}
\mathcal{W}\left(\hat{\Phi}(x),\mu(x),\nu(x)
\right)=\int\mathcal{D}k e^{-i\int d^{d}x k(x) \hat{\Phi(x)}}\chi(k)
\end{equation}
which satisfies the following evolution equation \cite{rosapl}

     $$\dot{\mathcal{W}}\left(\Phi(x),\mu(x),\nu(x),t\right)=\left\{ \int d^{d} x\left[ \mu(x)
     \frac{\delta}{\delta\nu(x)} + 2\nu(x)\frac{\delta}{\delta\Phi(x)}
     \Delta\left[\left(\frac{\delta}{\delta\Phi(x)}\right)^{-1}\frac{\delta}{\delta\mu(x)}\right]\right.\right.$$

$$ + \frac{i}{\hbar} \left[ U\left[ \left( \frac{-\delta}{\delta \Phi(x)}\right)^{-1}\frac{\delta}{\delta\mu(x)}
-\frac{i\nu(x)\hbar}{2}\frac{\delta}{\delta\Phi(x)}\right]\right]$$

$$ \left.\left. -U\left[ \left( \frac{-\delta}{\delta \Phi(x)}\right)^{-1}\frac{\delta}{\delta\mu(x)}
+\frac{i\nu(x)\hbar}{2}\frac{\delta}{\delta\Phi(x)}\right]
\right]\right\}$$

\begin{equation}\label{evolequa}
\times \mathcal{W}\left(\Phi(x),\mu(x),\nu(x),t\right),
\end{equation}
where  $\Delta f(x)=f(x+\Delta x)-f(x)$ and the operator $\left(
-\delta/\delta \Phi(x)\right)^{-1}$ is defined by

$$\left( \frac{-\delta}{\delta \Phi(x)}\right)^{-1}
\int\mathcal{D}k e^{-i\int d^{d}x k(x)
\hat{\Phi(x)}}=\int\mathcal{D}k \frac{i}{k(x)}e^{-i\int d^{d}x
k(x) \hat{\Phi(x)}}\,,$$ and the dot represents the time
derivative. The above formulas allow to develop a theory of
cosmological perturbations at classical and quantum levels.

\section{Classical and Quantum cosmological perturbations}

In view of Hamiltonian (\ref{hamiltonian}) and equation
(\ref{evolequa}) the evolution of cosmological quantum
perturbations can be described in terms of a marginal distribution
function  which satisfies the equation
$$\dot{\mathcal{W}}\left(v(x),\mu(x),\nu(x),t\right)=\left\{ \int d^{3} x\left[ \mu(x)
     \frac{\delta}{\delta\nu(x)} + 2\nu(x)\frac{\delta}{\delta v(x)}
     \Delta\left[\left(\frac{\delta}{\delta v(x)}\right)^{-1}\frac{\delta}{\delta\mu(x)}\right]\right.\right.$$
\begin{equation}\label{evolequa1}
\left.\left. -\frac{z''}{z}\left( 2
\nu(x)\frac{\delta}{\delta\mu(x)}\right)\right]\right\}\mathcal{W}\left(\hat{v}(x),\mu(x),\nu(x),t\right)\,.
\end{equation}
Due to the fact that the potential is quadratic, classical
perturbations follow a similar evolution equation. The only
difference is in the initial conditions which are restricted by
the Heisenberg uncertainty principle for the quantum
perturbations. A relevant expression is also the evolution
equation for the Fourier transform of the tomogram
\begin{equation}\label{FTT}
     \chi (k,\mu,\nu,t)=\int dX
     e^{ikX}{\mathcal{W}}\left(X,\mu,\nu,t\right)\,.
\end{equation}
It is

$$\dot{ \chi} (k(x),\mu(x),\nu(x),t) =\left\{\int d^{d}x
\left[\frac{1}{m}\mu(x)\frac{\partial}{\partial\nu(x)}-2i\hbar
k(x)\nu(x) \Delta\left[ \frac{1}{i
k(x)}\frac{\delta}{\delta\mu(x)}\right]\right] \right.$$
\begin{equation} \label{evolutionftt}
\left. -2\frac{z''}{z}\nu(x)\frac{\delta}{\delta\mu(x)}\right\}
\chi (k(x),\mu(x),\nu(x),t)\,.
\end{equation}

\section{The one-mode-parametric quantum oscillator}

In the previous section, the Hamiltonian describing the
perturbations has been constructed and presented in the form of a
sum of Hamiltonians of oscillators with time-dependent
frequencies. In this section, we are going to consider in detail
the one-mode evolution for the quantum parametric oscillator both
in the Schr\"{o}dinger representation and in the tomographic
probability representation. Let us use, for the one-dimensional
parametric oscillator, dimensionless units, i.e. with the Planck
constant $\hbar=1$, the ``mass'' of the oscillator $m=1$, and the
time-dependent frequency, at the characteristic initial time
$t_{0}$,  equal to unity, i.e. $\omega(t_{0})=1$. The Hamiltonian
of the parametric oscillator in these units reads
\begin{equation}\label{4.1}
     \hat{H}=\frac{\hat{p}^{2}}{2}+\frac{\omega^{2}(t)\hat{q}^{2}}{2}\,.
\end{equation}
This system has two integrals of motion, linear in position and
momentum \cite{pl1970,markov},
\begin{equation}\label{4.2}
  \hat{ A}(t)=\frac{i}{\sqrt{2}}\left( \varepsilon(t) \hat{p}-\dot{\varepsilon}(t)\hat{q}\right),
\end{equation}
\begin{equation}\label{4.3}
     \hat{A}^{\dag}(t)=-\frac{i}{\sqrt{2}}\left( \varepsilon^{*}(t) \hat{p}-\dot{\varepsilon}^{*}(t)\hat{q}\right).
\end{equation}
In Eq.(\ref{4.2}), the complex function of  time $\varepsilon(t)$
(where $t$  can be the conformal time) obeys the classical
equation of motion for the oscillator
\begin{equation}\label{4.4}
    \ddot{\varepsilon}(t) + \omega^{2}(t)\varepsilon(t)=0\,.
\end{equation}
The initial conditions
\begin{equation}\label{4.5}
    \varepsilon(t_{0})=1\,, \ \ \ \ \ \ \ \ \ \ \ \ \ \ \
    \dot{\varepsilon}(t_{0})=i\,,
\end{equation}
provide the commutation relations of the integrals of motion (\ref{4.2}) and (\ref{4.3})
\begin{equation}\label{4.6}
    [\hat{A}(t),\hat{A}^{\dag}(t)]=1\,.
\end{equation}
For the initial time $t_{0}$  (we shall use the initial time
$t_{0}=0$) the integrals of motion coincide with the standard
creation and annihilation operators
\begin{equation}\label{4.7}
    \hat{A}(t_{0})=a \,, \ \ \ \ \ \ \ \ \ \ \ \ \ \ \
    \hat{A}^{\dag}(t_{0})=a^{\dag}\,,
\end{equation}
where
\begin{equation}\label{4.8}
     \hat{a}=\frac{1}{\sqrt{2}}\left(\hat{q}+i\hat{p}\right)\,, \ \ \ \ \ \ \ \ \ \ \ \ \ \
     \hat{a}^{\dag}=\frac{1}{\sqrt{2}}\left(\hat{q}-i\hat{p}\right)\,.
\end{equation}
Here $\hat{q}$ and $\hat{p}$ are the position and momentum operators, respectively. The Schr\"{o}dinger equation
\begin{equation}\label{4.9}
    i\dot{\psi}(x,t)=-\frac{1}{2}\frac{\partial^{\,2}\psi(x,t)}{\partial
    x^{2}}+\frac{\omega^{2}(t)x^{2}}{2}\psi(x,t)\,,
\end{equation}
has the solution $\psi_{0}(x,t)$ which corresponds to the initial vacuum state, obeying the equation
\begin{equation}\label{4.10}
    \hat{A}(t)\psi_{0}(x,t)=0\,,
\end{equation}
with the initial condition
\begin{equation}\label{4.11}
    \psi_{0}(x,t_{0})=\frac{1}{\sqrt[4]{\pi}}e^{- x^{2} /2}\,,
\end{equation}
which is the standard ground state of the oscillator obeying the
vacuum condition
\begin{equation}\label{4.12}
    \hat{a}\psi_{0}(x,t_{0})=0\,.
\end{equation}
The solution $\psi_{0}(x,t)$ has the form of a Gaussian wave-packet
\begin{equation}\label{4.13}
    \psi_{0}(x,t)=\frac{1}{\sqrt[4]{\pi}}\frac{1}{\sqrt{\varepsilon(t)}}
    e^{\frac{i\dot{\varepsilon}((t)x^{2}}{2\varepsilon(t)}}\,.
\end{equation}
The Fock states $\psi_{n}(x,t)$, which are solutions  of the
Schr\"{o}dinger  Eq.(\ref{4.9}), are constructed by starting from
the vacuum state (\ref{4.13}) by standard algebraic formulas using
the integrals of motion (\ref{4.2}) and (\ref{4.3})
\begin{equation}\label{4.14}
     \psi_{n}(x,t)=\frac{1}{\sqrt{n!}}(A^{\dag}(t))^{N}\psi_{0}(x,t)\,.
\end{equation}
Solutions (\ref{4.14}) have the explicit form
\begin{equation}\label{4.15}
   \psi_{n}(x,t)=\psi_{0}(x,t)\frac{1}{\sqrt{2^{n}n!}}
   \left(\frac{\dot{\varepsilon}(t)}{\varepsilon(t)}\right)^{n/2}
    H_{n}\left(\frac{x}{|\varepsilon(t)|^{2}}\right)\,.
\end{equation}
Here $H_{n}$ are Hermite's polynomials. There exist Gaussian
packets which are squeezed coherent states which are obtained by
means of the Weyl displacement operator acting on the vacuum
state, i.e.
\begin{equation}\label{4.16}
  \psi_{\alpha}(x,t)=\hat{\mathcal{D}}(\alpha)\psi_{0}(x,t)\,,
\end{equation}
where the Weyl system reads
\begin{equation}\label{4.17}
 \hat{\mathcal{D}}(\alpha)=\exp\left( \alpha \hat{A}^{\dag}(t)-
 \alpha^{*}\hat{A}(t)\right)\,.
\end{equation}
Here $\alpha$ are complex numbers $\alpha=\alpha_{1} +
i\alpha_{2}$ and the squeezed coherent states have the properties
\begin{equation}\label{4.18}
    \int \psi_{\alpha}^{*}(x,t) \psi_{\beta}(x,t)dx=\exp \left(-\frac{|\alpha|^{2}}{2}-
    \frac{|\beta|^{2}}{2}+\alpha^{*}\beta\right)\,,
\end{equation}
and
\begin{equation}\label{4.19}
\frac{1}{\pi}\int\int \psi_{\alpha}^{*}(x,t)
\psi_{\alpha}(x',t)d\alpha_{1}\alpha_{2}=\delta(x-x')\,.
\end{equation}
Let us now consider the  tomographic probability description of
the quantum parametric oscillator. The symplectic tomogram of the
oscillator quantum vacuum state $\psi_{0}(x,t)$ is expressed in
terms of the wave function
\begin{equation}\label{4.20}
     \mathcal{W}_{0}(x,\mu,\nu,t)=\frac{1}{2\pi|\nu|}\left|\int\psi_{0}(y,t)e^{i\mu \frac{y^{2}}{2\nu}-i\frac{Xy}{\nu}}dy\right|^{2}.
\end{equation}
Using the explicit expression of the wave function, we get a
Gaussian tomographic probability distribution of the form
\begin{equation}\label{4.21}
   \mathcal{W}_{0}(x,\mu,\nu,t)=\frac{1}{\sqrt{2\pi \sigma^{2}_{\mu\nu}}}
    \exp\left( - \frac{X^{2}}{2\sigma^{2}_{\mu\nu}}\right)\,,
\end{equation}
where the dispersion of the  random position $X$ depends on the
parameters $\mu$ and $\nu$ and the function $\varepsilon(t)$ as
follows
\begin{equation}\label{4.22}
    \sigma^{2}_{\mu\nu} =\mu^{2}\sigma_{qq}+\nu^{2}\sigma_{pp}+2\mu\nu\sigma_{pq}
\end{equation}
where
\begin{equation}\label{4.23}
    \sigma_{qq}=\frac{|\varepsilon(t)|^{2}}{2}\ \ \ \ \ \ \ \sigma_{pp}=\frac{|\dot{\varepsilon}(t)|^{2}}{2}
    \ \ \ \ \ \ \
    \sigma^{2}_{pq}=\frac{1}{4}\left(|\varepsilon(t)\dot{\varepsilon(t)}|^{2}-1\right)\,.
\end{equation}
The state corresponding to   the tomogram (\ref{4.21}) is the
squeezed vacuum state. Depending on $\varepsilon(t)$ and
$\dot{\varepsilon}(t)$ the fluctuations of position or momentum
can be either smaller than  $1/2$ or larger than  $1/2$. The state
has the position-momentum correlation $\sigma_{pq}\neq 0$ and this
correlation satisfies the minimization of the
Schr\"{o}dinger-Robertson uncertainty relation \cite{sch,rob}
\begin{equation}\label{4.24}
     \sigma_{pp}\sigma_{qq}-\sigma_{pq}^{2}\geq\frac{1}{4}\,.
\end{equation}
The coherent state $\psi_{\alpha}(x,t)$ has also a Gaussian tomogram
\begin{equation}\label{4.25}
\mathcal{W}_{\alpha}(X,\mu,\nu,t)=\frac{1}{\sqrt{2\pi
\sigma^{2}_{\mu\nu}}}\exp \left[-\frac{\left(X-\bar{X}\right)}{2
\sigma^{2}_{\mu\nu}}\right]\,,
\end{equation}
where the dispersion $\sigma^{2}_{\mu\nu}$ is given by Eqs.
(\ref{4.22}) and (\ref{4.23}) and the mean value reads
\begin{equation}\label{4.26}
    \bar{X}=\mu \langle \hat{q}\rangle_{\alpha}+ \nu
    \langle\hat{p}\rangle_{\alpha}\,,
\end{equation}
where
\begin{equation}\label{4.27}
\langle \hat{q}\rangle_{\alpha}=\sqrt{2}\,\rm{Re}\, \alpha(t)\,, \
\ \ \ \ \ \ \ \  \langle
\hat{p}\rangle_{\alpha}=\sqrt{2}\,\rm{Im}\, \alpha(t)\,.
\end{equation}
Here
\begin{eqnarray}
\nonumber \alpha(t) &=& \frac{i}{\sqrt{2}}\left(\varepsilon(t)\bar{p}- \dot{\varepsilon(t)}\bar{q} \right)\,, \\
    \alpha^{*}(t) &=& -\frac{i}{\sqrt{2}}\left(\varepsilon(t)\bar{p}- \dot{\varepsilon(t)}\bar{q}
    \right)\,,
     \end{eqnarray}
and
\begin{equation}\label{4.29}
\bar{q}=\sqrt{2}\,\rm{Re}\, \alpha\,, \ \ \ \ \ \ \ \ \
\bar{p}=\sqrt{2}\,\rm{Im}\, \alpha\,,
\end{equation}
where $\alpha=\alpha_{1}+i\alpha_{2}$, i.e. $\alpha$ is a constant
complex number labelling coherent states $\langle
x|\alpha,t\rangle=\psi_{\alpha}(x,t)$.

The probability distributions $\mathcal{W}_{0}(X,\mu,\nu,t)$ and
$\mathcal{W}_{\alpha}(X,\mu,\nu,t)$ satisfy the kinetic equation
\begin{equation}\label{4.30}
    \dot{\mathcal{W}}(X,\mu,\nu,t)-\mu\frac{\partial}{\partial\nu}\mathcal{W}(X,\mu,\nu,t)+
    \omega^{2}(t)\nu\frac{\partial}{\partial\mu}\mathcal{W}(X,\mu,\nu,t)=0\,.
\end{equation}

The kinetic equation corresponds to the classical Liouville
equation in classical mechanics, that is
\begin{equation}\label{4.31}
    \frac{\partial f(q,p,t)}{\partial t} + \frac{\partial f(q,p,t)}{\partial q}p- \frac{\partial f(q,p,t)}{\partial p}
     \cdot\frac{\partial V(q,t)}{\partial q}=0\,,
\end{equation}
where the potential energy $V(q,t)$ is the energy for the parametric oscillator
\begin{equation}\label{4.32}
     V(q,t)=\frac{1}{2} \omega^{2}(t) q^{2}\,.
\end{equation}
However, the same Eq.(\ref{4.30}) corresponds to the quantum von
Neumann equation for the density operator of the parametric
oscillator
\begin{equation}\label{4.33}
    i \frac{\partial }{\partial t}\hat{\rho}(t)=[\hat{H}(t),
    \hat{\rho(t)}]\,,
\end{equation}
being $\hbar=1$, where
\begin{equation}\label{4.34}
    \hat{H}(t)=\frac{\hat{p}^{2}}{2}+\frac{\omega^{2}(t) \hat{q}^{2}}{2}\,.
\end{equation}
In the tomographic probability representation Eqs. (\ref{4.31})
and (\ref{4.33}) coincide. The evolution of the tomogram
$\mathcal{W}(X,\mu,\nu,t)$ can be expressed in terms of the
propagator
\begin{equation}\label{4.35}
    \mathcal{W}(X,\mu,\nu,t)=\int\Pi(X,\mu,\nu,X',\mu',\nu',t)\mathcal{W}(X',\mu',\nu',0)dX'd\mu'd\nu'\,.
\end{equation}
The propagator  $\Pi(X,\mu,\nu,X',\mu',\nu',t)$ in this integral
relation is identical for both classical and quantum  evolutions
of the one-mode-parametric oscillator with a time-dependent
frequency. The difference appears in the initial conditions.
However, classical tomograms have  not to respect the uncertainty
relation
$$\left\{\int X^{2} \mathcal{W}(X,\mu,\nu,t)dX|_{\mu=1,\nu=0}-\left[\int X \mathcal{W}(X,\mu,\nu,t)\right]^{2}_{\mu=1,\nu=0}\right\} $$
\begin{equation}\label{4.36}
     \times \left\{\int X^{2} \mathcal{W}(X,\mu,\nu,t)dX|_{\mu=0,\nu=1}-
     \left[\int X
     \mathcal{W}(X,\mu,\nu,t)\right]^{2}_{\mu=0,\nu=1}\right\}\geq\frac{1}{4}\,,
\end{equation}
while the tomogram of a quantum oscillator must satisfy these
inequalities. This results means that,  since the evolution is
governed by the same equations, both situations, classical and
quantum, are represented. Since the field is the collection of
modes, the arguments presented above are applicable to all the
fields. This means that for the field tomogram, which is the
product of tomograms of all modes, the evolution result classical.
{\it Quantumness} is hidden in the initial state but summing up
all  states gives, as result,  the classical field. The
perturbations described by the Hamiltonian, thus in the
tomographic picture, can be discussed both using the quantum or
the classical propagator of the field since both coincide.

It is straightforward that the Hamiltonian describing the
cosmological perturbations is the Hamiltonian of field harmonic
vibrations. It is the natural Hamiltonian of a system whose
properties fluctuate around the background state (the FLRW
background) and where the fluctuations are small (small
vibrations). The deviations from the ``equilibrium'' background
states are obeying the Hooke law. Only for  large deviations from
the equilibrium state, the Hooke law is violated and strong
anharmonicity  appears. The ``spring'' providing the harmonic
vibrations of a system with very large amplitude of vibrations can
even break giving possibility to the system to come out far away.
This simple mechanism in the classical description of small
vibrations and in quantum domain can give different results. In
fact the ``ground'' state of the field is the squeezed state of a
harmonic linear oscillators with time-dependent frequencies. The
fluctuations of these vibrations in the classical domain, at zero
temperature, are also equal to zero. The small vibrations cannot
provide, in this case, large deviations from the equilibrium state
even if one takes into account the possible appearance of Hooke's
law violation for large amplitudes of the oscillations. For a
quantum domain, even these small vibrations have quantum
fluctuations. Thus the deviations of the oscillation position
$|\varepsilon(t)|^{2}/2$ and of the momentum
$|\dot{\varepsilon}(t)|^{2}/2$ can become large in comparison with
the standard vacuum ones and be equal to $1/2$, due to the
influence of the possible large values of the contributions of
$|\varepsilon(t)|$ (or $|\dot{\varepsilon}(t)|$). This mechanism
of quantum fluctuations in presence of  time-dependent frequencies
can, for some parts of the system, create large amplitudes which
violate the  Hooke  law. In such a case, the system can give rise
to domains of larger densities and non-uniformities which results
as inhomogeneities and anisotropies with respect to the
background. This could be a coherent scheme to match quantum
microscopic primordial perturbations with the today observed large
scale structure.

\section{The multi-mode parametric small perturbations}

Let us consider now the most general multi-mode Hamiltonian. This
situation results more realistic in order to take into account the
problem of cosmological perturbations. As in \cite{Kiefer:1998qe},
we consider periodic conditions. Due to these, the Hamiltonian can
be taken in the following form
\begin{equation}\label{7.1}
    \mathcal{H}(t)=\frac{1}{2}\bar{}\vec{Q}B(t)\vec{Q}+
    \vec{C}(t)\vec{Q}\,.
\end{equation}
Here we consider $N$ modes ($N$ can be equal to $\infty$) and the vector operator
\begin{equation}\label{7.2}
\vec{Q}=\left( \hat{P}_{1},\hat{P}_{2} ,\dots,\hat{P}_{N},
\hat{q}_{1}, \hat{q}_{2},\dots, \hat{q}_{N}\right)\,.
\end{equation}
The system properties are coded by the interaction $2N\times 2N$
matrix $B(t)$ which depends on time $t$,that is
\begin{equation}\label{7.3}
     B(t)=\left(
            \begin{array}{cc}
              B_{1}(t) & B_{2}(t) \\
              B_{3}(t) &  B_{4}(t) \\
            \end{array}
          \right)\,.
\end{equation}
The $N\times N$ block matrices $B_{k}(t)$, $k=1,2,3,4$ correspond
to quadratures interactions, namely $B_{1}(t)$ describes
$\hat{P}_{k}-$quadratures interactions, $B_{4}(t)$ corresponds to
$\hat{q}_{k}-$quadratures interactions and the matrices $B_{2}(t)$
and $B_{3}(t)$ to interaction terms due to $\hat{q}$ and $\hat{P}$
couplings. The  $2N$-vector $\vec{C}(t)$ provides the interaction
terms corresponding to homogeneous "electric-like" fields acting
on charged particles.

The classical counterpart of this Hamiltonian, in the same
quadratic form,  is given by a vector (\ref{7.2}) composed by
classical momenta and positions with standard Poisson brackets.
The solution of the Schr\"{o}dinger equation with first integrals
of motion, in position and momenta, determines an inhomogeneous
symplectic group element parameterized by $2N\times 2N$ matrix
$\Lambda(t)$ and a  $2N-$vector $\vec{\Delta}(t)$. One has the
$2N-$vector $\vec{I}(t)$ whose components are integrals of motion.
The vector reads as
\begin{equation}\label{7.4}
     \vec{I}(t)=\Lambda(t) \vec{Q}+\vec{\Delta}(t).
\end{equation}
The symplectic matrix  $\Lambda(t)$ satisfies the equation
\begin{equation}\label{7.5}
     \dot{\Lambda}(t)=\Lambda(t)\Sigma B(t),\ \ \ \ \ \ \ \     \Sigma=\left(
                                                                         \begin{array}{cc}
                                                                           0 & -1_{N} \\
                                                                            1_{N} & 0 \\
                                                                         \end{array}
                                                                       \right)
\end{equation}
and the initial condition
\begin{equation}\label{7.6}
    \Lambda(0)=1_{2N}\,.
\end{equation}
The vector $\vec{\Delta}(t)$ satisfies the evolution equation
\begin{equation}\label{7.7}
    \dot{\vec{\Delta}}=\Lambda(t)\Sigma\vec{C}(t)
\end{equation}
with initial  value, $\Delta(0)=0$\,.

The evolution of the system of field modes with  Hamiltonian
(\ref{7.1}) is described by the  Green function of the
Schr\"{o}dinger non-stationary equation determined in terms of the
matrix $\Lambda(t)$, the vector $\vec{\Delta}$, and the Gaussian
Green function \cite{markov}. It is
$$G(\vec{x},\vec{x}',t)=[det[-2\pi_{i}\lambda_{3}]^{-1/2}\exp\left\{-\frac{i}{2}\left[\vec{x} \lambda^{-1}_{3}\lambda_{1} \vec{x}-2\vec{x}\lambda^{-1}_{3}\vec{x}'+ \vec{x}'\lambda_{1}\lambda_{3}^{-1}\vec{x}'
\phantom{\int} \right.\right.$$
\begin{equation}\label{7.8}
  \left. \left.+2\vec{x}'  \lambda^{-1}_{3}\vec{\delta}_{2}+2\vec{x}(\vec{\delta}_{1}-\lambda_{1}\lambda_{3}^{-1}\vec{\delta}_{2}) +\vec{\delta}_{2}\lambda_{1}\lambda_{3}^{-1}\vec{\delta}_{2}
    -2\int_{0}^{t}\dot{\vec{\delta}_{1}}(\tau)\vec{\delta}_{2}(\tau) d\tau \right]
    \right\}\,.
\end{equation}
Here, the symplectic matrix $\Lambda(t)$ as well as the vector
$\vec{\Delta}$ are  in a block form being
\begin{equation}\label{7.9}
    \Lambda(t)=\left(
                 \begin{array}{cc}
                   \lambda_{1}(t) &  \lambda_{2}(t) \\
                    \lambda_{3}(t) &  \lambda_{4}(t) \\
                 \end{array}
               \right);     \ \ \ \ \ \
               \vec{\Delta}(t)=\left(
                                 \begin{array}{c}
                                   \vec{\delta}_{1}(t) \\
                                   \vec{\delta}_{2}(t) \\
                                 \end{array}
                               \right)\,.
\end{equation}
This means that, given  the initial quantum state at $t=0$,
$\psi(\vec{x},0)$ where $\vec{x}=(x_{1},x_{2},\dots x_{N})$, the
state at time $t$ reads
\begin{equation}\label{7.10}
    \psi(\vec{x},t)=\int G(\vec{x},\vec{x}',t)\psi(\vec{x},t)dx'\,.
\end{equation}
If the initial state is given by a density matrix
$\rho(\vec{x},\vec{x}',0)$,  the density matrix at time $t$ reads
\begin{equation}\label{7.11}
    \rho(\vec{x},\vec{x}',t)=\int G(\vec{x},\vec{y},t)\rho(\vec{y},\vec{y}',0)G^{*}(\vec{x}',\vec{y}',t)
    d\vec{y}d\vec{y}'\,.
\end{equation}
In \cite{Lesgourgues:1996jc}, some initial states are studied. The
first is that the initial state is the vacuum state with a
Gaussian wave function corresponding to a set of oscillators with
different frequencies. Another possibility is that the initial
state is chosen  as a  thermal density matrix of the set of
oscillators at  temperature $T$. The  density matrix is also
Gaussian and the state at $t=0$ can be chosen  as a pure squeezed
state with a large, but finite, number of particles with wave
functions expressed in terms of Hermite polynomials. The true
vacuum state and the thermal  state can be considered in both
classical and quantum universe evolution pictures. The squeezed
Fock states of the evolving Universe correspond to the quantum
domain  only because the states cannot be realized as initial
state in the classical domain. Below we focus on the vacuum
initial state to show, in tomographic probability representation,
its evolution both in classical and in quantum pictures.

The quantum  tomogram
$\mathcal{W}(\vec{X},\vec{\mu},\vec{\nu},t)$, where
$\vec{X}=(X_{1}, X_{2},\dots, X_{N})$, $\vec{\mu}=(\mu_{1},
\mu_{2},\dots, \mu_{N})$, $\vec{\nu}=(\nu_{1}, \nu_{2},\dots,
\nu_{N})$ which is a joint probability density of random variables
$X_{k}$, $k=1,2,\dots, N$ is expressed in terms of the wave
function
\begin{equation}\label{7.12}
\mathcal{W}(\vec{X},\vec{\mu},\vec{\nu},t)=
\frac{1}{\prod_{k=1}^{N}2\pi|\nu_{k} |} \left|
\int\psi(\vec{y},t)\exp\left\{i\left[\sum_{k=1}^{N}\left(\frac{\mu_{k}}{2\nu_{k}}y^{2}_{k}
-\frac{X_{k}y_{k}}{\nu_{k}}
\right)\right]\right\}d\vec{y}\right|^{2}\,.
\end{equation}
The Wigner function of the state is expressed as the  wave
function
\begin{equation}\label{7.13}
     W(\vec{q},\vec{p},t)=\int\psi\left(\vec{q}+\frac{\vec{u}}{2},t\right)\psi^{*}\left(\vec{q}-\frac{\vec{u}}{2},t\right)
     e^{-i\vec{p}\cdot\vec{u}}d\vec{u}\,.
\end{equation}
The tomogram (\ref{7.12}) is connected with the above Wigner function by the Radon transform
\begin{equation}\label{7.14}
     \mathcal{W}(\vec{X},\vec{\mu},\vec{\nu},t)=\int W(\vec{q},\vec{p},t)
     \prod_{k=1}^{N}\left[\delta\left(X_{k}-\mu_{k}q_{k}-\nu_{k}p_{k}\right)\frac{dq_{k}dp_{k}}{2\pi}\right]\,.
\end{equation}
The density matrix in position representation
satisfies the von Neumann equation
\begin{eqnarray}
  \nonumber i\frac{\partial \rho (\vec{x},\vec{x'},t)}{\partial t} &=& \frac{1}{2}\left(\vec{Q}_{\vec{x}}B(t)\vec{Q}_{\vec{x}}\right)\rho (\vec{x},\vec{x'},t)
+\vec{C}(t)\vec{Q}_{\vec{x}}\,\rho (\vec{x},\vec{x'},t) \\
  & &  -\frac{1}{2}\left(\vec{Q}_{\vec{x'}}B(t)\vec{Q}_{\vec{x'}}\right)\rho (\vec{x},\vec{x'},t)
-\vec{C}(t)\vec{Q}_{\vec{x'}}\,\rho (\vec{x},\vec{x'},t)\,.\ \ \
\label{7.15}
\end{eqnarray}
Here the operators $\vec{Q}_{\vec{x}}$ and $\vec{Q}_{\vec{x'}}$
are given by (\ref{7.2})  where $\hat{P}_{k \vec{x}}\rightarrow
-i\frac{\partial}{\partial x_{k}}$, $\hat{q}_{k
\vec{x}}\rightarrow x_{k}$ and $\hat{P}_{k \vec{x'}}\rightarrow
-i\frac{\partial}{\partial x'_{k}}$, $\hat{q}_{k
\vec{x'}}\rightarrow x'_{k}$ respectively. The equation for the
Wigner function (the Moyal equation, see \cite{moyal49}) is given
by (\ref{7.15}) with the replacements
\begin{eqnarray}
 \nonumber  \frac{\partial}{\partial x_{k}} \rightarrow \frac{1}{2}\frac{\partial}{\partial q_{k}}+i p_{k},& & \frac{\partial}{\partial x'_{k}} \rightarrow \frac{1}{2}\frac{\partial}{\partial q_{k}}-i p_{k}  \\
  x_{k}\rightarrow q_{k}+ \frac{i}{2}\frac{\partial}{\partial p_{k}},& & x'_{k}\rightarrow q_{k}
  - \frac{i}{2}\frac{\partial}{\partial p_{k}}\, .\label{7.16}
\end{eqnarray}

The evolution equation for the tomogram (\ref{7.12}) or (\ref{7.14}) is obtained from the equation for the
Wigner function by the replacements
\begin{eqnarray}
 \nonumber  W(\vec{q},\vec{p},t) &\rightarrow & \mathcal{W}(\vec{X},\vec{\mu},\vec{\nu},t)  \\
 \nonumber   \frac{\partial}{\partial q_{k}}\rightarrow  \mu\frac{\partial}{\partial X_{k}}& ;&\frac{\partial}{\partial P_{k}}\rightarrow  \nu\frac{\partial}{\partial X_{k}}   \\
   q_{k}\rightarrow -\frac{\partial}{\partial \mu_{k}}\left(\frac{\partial}{\partial X_{k}}\right)^{-1}  &;&
   P_{k}\rightarrow -\frac{\partial}{\partial \nu_{k}}\left(\frac{\partial}{\partial
   X_{k}}\right)^{-1}\,.
   \label{7.17a}
\end{eqnarray}

It is important to point out that the equation for the Wigner
function and for the tomogram will coincide for both domains
quantum and classical  with the equations for probability
densities $f(\vec{q},\vec{p},t)$ satisfying the Liouville equation
and for the quantum and the classical tomogram
\begin{equation}\label{7.17b}
     \mathcal{W}(\vec{X},\vec{\mu},\vec{\nu},t)=\int f(\vec{q},\vec{p},t)\prod_{k=1}^{N}
     \left[\delta\left(X_{k}-\mu_{k}q_{k}-\nu_{k}p_{k}\right)
     dq_{k}dp_{k}\right]\,,
\end{equation}
respectively.

The tomogram is the probability density.  Due to this one can
introduce the Shannon entropy \cite{Man'ko:2006wv} associated with
the probability density. The formula for the entropy reads
\begin{equation}\label{7.17c}
    H_{1}(\vec{\theta})=-\int\mathcal{W}(\vec{X},\vec{\mu},\vec{\nu})\ln\mathcal{W}(\vec{X},\vec{\mu},\vec{\nu})d\vec{x}\,,
\end{equation}
$$\mu_{k}=s_{k}\cos \theta_{k},\ \ \ \ \ \  \nu_{k}=s^{-1}_{k}\sin\theta_{k} \,.$$
It means that for evolution  of the Wigner function $
W(\vec{q},\vec{p},t)$, in the quantum domain, and the probability
distribution $ f(\vec{q},\vec{p},t)$, in the classical domain, the
propagators are identical as well  as propagators for tomograms
both in quantum and classical domains. It is a property of systems
with Hamiltonians of the form (\ref{7.1}) and it corresponds to
the Ehrenfest theorem.

Thus, given the evolution of the tomogram in two states, the
evolution of entropy can be defined as
\begin{equation}
 H(\vec{\theta}, t)=H_{1}(\vec{\theta}, t)+H_{2}(\vec{\theta},
t)\,.
\end{equation}
where
\begin{equation}\label{7.100}
     \vec{\theta}= (\theta_{1},\theta_{2},\dots \theta_{N})\,,
\end{equation}
and \begin{equation}
H_{2}(\vec{\theta})=H_{1}\left(\vec{\theta}+\frac{\vec{\pi}}{2}\right),\
\ \ \ \ \ \ \vec{\theta}+\frac{\vec{\pi}}{2}= \left\{\theta_{k}
+\frac{ \pi }{2} \right\}. \end{equation}
One can introduce the sum entropy
\begin{equation}\label{7.101}
     \tilde{H}(\vec{\theta},t)= H(\vec{\theta},t)+ H\left(\vec{\theta}+ \frac{\vec{\pi}}{2},t\right)\,,
\end{equation}
where
\begin{equation}\label{7.102}
 \vec{\theta}+ \frac{\vec{\pi}}{2}=\left\{ \theta_{k}+
 \frac{\pi}{2}\right\}\,.
\end{equation}
The entropy $\tilde{H}(\vec{\theta},t) $ satisfies the inequality, see e.g.\cite{Man'ko:2006wv},
\begin{equation}\label{7.103}
    \tilde{H}(\vec{\theta},t)\geq N \ln \pi e\,.
\end{equation}
For the initial ground state of the universe, one has the
saturation of the above inequality, i.e.
\begin{equation}\label{7.103a}
    \tilde{H}(\vec{\theta} )= N \ln \pi e\,.
\end{equation}
In the quantum domain, the entropy $ \tilde{H}(\vec{\theta} )$
cannot be less than the value $N \ln \pi e$. In the classical
domain, the entropy $ \tilde{H}(\vec{\theta} )$ characterizes the
order or disorder in the field state. In the process of evolution
of the Universe, this entropy is changing. The tomograms for
classical states  are probability distributions evolving for the
quadratic Hamiltonian (\ref{7.1}) identically. If one measures the
tomogram at the late time $\bar{t}$, corresponding to the
classical epoch, one can learn what was the initial state since at
this period of time the tomogram  corresponding to the classical
state provides the possibility to calculate the probability
density $f(\vec{q},\vec{p},t)$, using the inverse Radon transform
\begin{equation}\label{R1}
     f(\vec{q},\vec{p},t) =\frac{1}{(2\pi)^{N}}\int \mathcal{W}(\vec{X},\vec{\mu},
     \vec{\nu},t)e^{i\sum_{k=1}^{N}(X_{k}-\mu_{k}q_{k}-\nu_{k}p_{k})}\prod_{k=1}^{N}dX_{k}d\mu_{k}d\nu_{k}\,.
\end{equation}
In principle, this density can be used to describe the
distribution of matter in galaxies \cite{binney} and in clusters
of galaxies \cite{peacock}. Besides, one can introduce another
characterization of the Universe state in terms of the tomogram.

\section{Conclusions}

To conclude, we point out the main results of the paper. We have
considered classical and quantum perturbations under the same
tomographic standard. The small vibration Hamiltonian with
time-dependent parameters can give account of the basic model
which can be adopted in the quantum and in the classical regime.
Specifically, the classical and quantum descriptions  can be
associated with the tomographic probability distributions. The
classical initial states (classical tomograms) and quantum initial
states (quantum tomograms) have different properties because the
classical tomographic entropies can violate the quantum bound of
the tomographic entropy. The tomographic entropy can be estimated
at present epoch by measuring the space and momentum distributions
of the matter in the universe. Since the propagator for both
classical and quantum tomograms of the universe states is the
same, one can trace back, in principle, the entropy data to the
initial state of the universe. The observed data, extrapolated to
the initial state by means of the classical tomogram, can be less
than the quantum bound. The existence of bound property can be
used to discriminate between cosmological quantum and classical
behaviors.

The physical  meaning of the  entropy which we have considered is
that it represents the sum of two entropies. One corresponds to
the spatial probability density of matter. Then, we have to assume
$\vec{\theta}=0$. The other one corresponds to the momentum
distribution. For this case, we have to assume the value
$\vec{\theta}=\frac{\vec{\pi}}{2}$. By measuring these
distributions at  present epoch, one can trace back the
cosmological evolution to entropy initial values  corresponding to
the small vibrations. In other words, the tomographic description
of the universe  provides the possibility to correlate classical
and quantum cosmological perturbations in the same unitary scheme.
The final goal of this picture is to find out a dynamical and
self-consistent approach capable of connecting the primordial
quantum perturbations to the today observed large scale structure
\cite{peacock}. In a forthcoming paper, we will discuss the above
results considering  the  data coming from the observations and
their relations with  initial values of cosmological parameters.

\section*{Acknowledgments}
V.I. Man'ko wants to thank the University of Naples "Federico II"
and the INFN, Sez. di Napoli, for the hospitality.


\begin{thebibliography}{99}
\bibitem{Guth:1980zm}
  A.~H.~Guth,
  Phys.\ Rev.\  D {\bf 23} (1981) 347.
\bibitem{Linde:1981mu}
  A.~D.~Linde,
  Phys.\ Lett.\  B {\bf 108} (1982) 389.

\bibitem{Mukhanov:1990me}
  V.~F.~Mukhanov, H.~A.~Feldman and R.~H.~Brandenberger,
  ``Theory Of Cosmological Perturbations. Part 1. Classical Perturbations. Part
 2. Quantum Theory Of Perturbations. Part 3. Extensions,''
  Phys.\ Rept.\  {\bf 215} (1992) 203.

\bibitem{Lesgourgues:1996jc}
  J.~Lesgourgues, D.~Polarski and A.~A.~Starobinsky,
  Nucl.\ Phys.\  B {\bf 497} (1997) 479
  [arXiv:gr-qc/9611019].

  \bibitem{Kiefer:1998qe}
  C.~Kiefer, D.~Polarski and A.~A.~Starobinsky,
  Int.\ J.\ Mod.\ Phys.\  D {\bf 7} (1998) 455
  [arXiv:gr-qc/9802003].


\bibitem{schroedinger26}
E. Schroedinger,    "An Undulatory Theory of the Mechanics of
Atoms and Molecules" (PDF). Phys. Rev. 28 (6)  1049–1070 (1926)


\bibitem{landau} L. D. Landau, Z. Physik, {\bf 45} 430 (1927)

\bibitem{vonneumann32} J. von Neumann ``Mathematische
Grundlagen der Quantenmechanik'', Springer Verlag, Berlin 1932

\bibitem {hl}
 J. J. Halliwell, Phys. Rev. D {\bf 38}, 2468, (1988)



\bibitem{Mancini:1996wd}
  S.~Mancini, V.~I.~Man'ko and P.~Tombesi,
  Phys.\ Lett.\  A {\bf 213} (1996) 1
  [arXiv:quant-ph/9603002].
\bibitem{9qm}
D. F. Styer et al.,
American Journal of Physics {\bf 70},   288-297 (2002)
\bibitem{radon} J. Radon, Ber. Verh. Sachs. Acad. {\bf 69}, 269
(1917)

\bibitem{olga97}O. V. Manko, V. I. Manko, J. Russ. Laser Res. {\bf 18},
 407 (1997)




\bibitem{Manko:2003dp} V.~I.~Manko, G.~Marmo and C.~Stornaiolo,
``Radon transform of Wheeler-De Witt equation and tomography of
quantum states of the universe,'' to appear on Gen.Rel.Grav.,
gr-qc/0307084.


\bibitem{Man'ko:2004zj}
  V.~I.~Man'ko, G.~Marmo and C.~Stornaiolo,
  Gen.\ Rel.\ Grav.\  {\bf 37} (2005) 2003
  [arXiv:gr-qc/0412091].

\bibitem{Man'ko:2006wv}
  V.~I.~Man'ko, G.~Marmo and C.~Stornaiolo,
  Gen.\ Rel.\ Grav.\  {\bf 40} (2008) 1449
  [arXiv:gr-qc/0612073].
\bibitem{Stornaiolo:2006da}
  C.~Stornaiolo,
  AIP Conf.\ Proc.\  {\bf 841} (2006) 645.


\bibitem{Stornaiolo:2006kz}
  C.~Stornaiolo,
  J.\ Phys.\ Conf.\ Ser.\  {\bf 33} (2006) 242.
\bibitem{Capozziello:2007xn}
  S.~Capozziello, V.~I.~Man'ko, G.~Marmo and C.~Stornaiolo,
  Gen.\ Rel.\ Grav.\  {\bf 40} (2008) 2627
  [arXiv:0706.3018 [gr-qc]].



\bibitem{rosapl} V. I. Manko, L. Rosa, P. Vitale
Phys. Lett B {\bf 439}, 328 (1998)

\bibitem{hus1953}K. Husimi, Progr. Theor. Phys. 9 (1953) 238

\bibitem{Malkin:1970gq}
  I.~A.~Malkin, V.~I.~Man'ko and D.~A.~Trifonov,
  Phys.\ Rev.\  D {\bf 2} (1970) 1371.

\bibitem{Malkin:1971gs}
  I.~A.~Malkin, V.~I.~Man'ko and D.~A.~Trifonov,
  J.\ Math.\ Phys.\  {\bf 14} (1973) 576.
  \bibitem{markov}  V. V. Dodonov, V.~I.~Man'ko  in ed. M. A. Markov,
   ``Invariants and the evolution of nonstationary quantum systems.''Moscow,
   Izdatel'stvo Nauka (AN SSSR, Fizicheskii Institut, Trudy. Volume 183), 1987, 288 p.
   In Russian. Translated into English, publisher Nova Science (1989).

\bibitem{copeland}
E.J. Copeland, M. Sami, S. Tsujikawa, Int. Jou. Mod. Phys. {\bf D
15}, 1753 (2006).

\bibitem{odintsov}
S. Nojiri and S.D. Odintsov, {\it Int. J. Meth. Mod. Phys.} {\bf
4}, 115 (2007).

\bibitem{faraoni}
T.P. Sotiriou  and  V. Faraoni, arXiv:0805.1726 [gr-qc], 2008.

\bibitem{GRGrew}
S. Capozziello and M. Francaviglia, Gen. Rel. Grav. {\bf 40}, 357
(2008).


\bibitem{DodDod}V. V. Dodonov, A. V. Dodonov, "Quantum harmonic oscillator and nonstationary Casimir effect",
Journal of Russian Laser Research 26, 445 (2005)

 \bibitem{pl1970}I. A. Malkin and V.1. Man'ko, Phys. Lett. A 32 (1970) 243.

\bibitem{moyal49}  J.~E.~Moyal, Proc. Cambridge Philos. Soc.
{\bf 45}, 99 (1949)

\bibitem{sch} E. Schr\"{o}dinger,   Sitzungsber, Preuss Akad. Wiss., p. 296 (1930).

\bibitem{rob} H.P. Robertson,  Phys. Rev. {\bf 35}, 667(A) (1930).

\bibitem{binney}
J. Binney, S. Tremaine, "Galactic dynamics", Princeton Univ.
Books, Princeton (1987).

\bibitem{peacock}
J.A. Peacock, "Cosmological Physics", Cambridge Univ. Press,
Cambridge (1999).




\end{thebibliography}
\end{document}